\documentclass[fleqn,usenatbib]{mnras}

\usepackage{newtxtext,newtxmath}
\usepackage[T1]{fontenc}
\usepackage{ae,aecompl}
\usepackage{hyperref}

\usepackage[usenames]{xcolor}
\usepackage{color, soul}

\usepackage{graphicx}	% Including figure files
\usepackage{amsmath}	% Advanced maths commands
\usepackage{amssymb}	% Extra maths symbols
\usepackage{times}
\title[Dust and Gas in IRS 48]{Signatures of an eccentric disc cavity: Dust and gas in IRS 48}
\author[J. Calcino et al.]{
Josh Calcino$^{1}$\thanks{Contact e-mail: \href{mailto:j.calcino@uq.edu.au}{j.calcino@uq.edu.au}},
Daniel J. Price$^{2}$,
Christophe Pinte$^{2,3}$,
Nienke van der Marel $^{4}$,
\newauthor{Enrico Ragusa$^5$, Giovanni Dipierro$^5$, Nicol\'as Cuello$^{6,7}$, Valentin Christiaens$^{2}$}
\\
$^{1}$School of Mathematics and Physics, The University of Queensland, QLD 4072, Australia\\
$^{2}$Monash Centre for Astrophysics (MoCA) and School of Physics and Astronomy, Monash University, Clayton Vic 3800, Australia\\
$^{3}$Univ. Grenoble Alpes, CNRS, IPAG, F-38000 Grenoble, France\\
$^{4}$Herzberg Astronomy \& Astrophysics Programs, National Research Council of Canada, 5071 West Saanich Road, Victoria BC V9E 2E7, Canada\\
$^{5}$Department of Physics and Astronomy, University of Leicester, Leicester, United Kingdom\\
$^{6}$Instituto de Astrof\'isica, Pontificia Universidad Cat\'olica de Chile, Santiago, Chile\\
$^{7}$N\'ucleo Milenio de Formaci\'on Planetaria (NPF), Chile\\
}
\date{Accepted XXX. Received YYY; in original form ZZZ}

\pubyear{2019}

\begin{document}
\label{firstpage}
\pagerange{\pageref{firstpage}--\pageref{lastpage}}
\maketitle

% Abstract of the paper
\begin{abstract}
We test the hypothesis that the disc cavity in the `transition disc' Oph IRS 48 is carved by an unseen binary companion. We use 3D dust-gas smoothed-particle hydrodynamics simulations to
demonstrate that marginally coupled dust grains concentrate in the gas over-density that forms in in the cavity around a low binary mass ratio binary. This produces high contrast ratio dust asymmetries at the cavity edge similar to those observed in the disc around IRS 48 and other transition discs. This structure was previously assumed to be a vortex. However, we show that the observed velocity map of IRS 48 displays a peculiar asymmetry that is not predicted by the vortex hypothesis. We show the unusual kinematics are naturally explained by the non-Keplerian flow of gas in an eccentric circumbinary cavity. We further show that perturbations observed in the isovelocity curves of IRS 48 may be explained as the product of the dynamical interaction between the companion and the disc. The presence of a $\sim$0.4 M$_{\odot}$ companion at a $\sim$10 au separation can qualitatively explain these observations. High spatial resolution line and continuum imaging should be able to confirm this hypothesis.
\end{abstract}

% Select between one and six entries from the list of approved keywords.
% Don't make up new ones.

%%%%%%%%%%%%%%%%%%%%%%%%%%%%%%%%%%%%%%%%%%%%%%%%%%

%%%%%%%%%%%%%%%%% BODY OF PAPER %%%%%%%%%%%%%%%%%%

\section{Introduction}

Protoplanetary discs with cleared out central cavities, i.e.~\emph{transition discs}, are promising sites for characterising planet formation. 
The cleared central regions may indicate the presence of companions \citep{strom1989,marsh1992,marsh1993}, but photoevaporation may also explain this feature \citep{alexander14a, turner2014} --- the latter hypothesis giving transition discs their name. 

Resolved observations of transition discs with the Atacama Large Millimetre/submillimetre Array (ALMA) have revealed large asymmetries in (sub-)millimeter continuum observations, with Oph IRS 48 \citep{vandermarel2013,vandermarel2015}, HD 142527 \citep{casassus2015}, and AB Aur \citep{tang2017} being a few notable examples. Dust asymmetries around the cavity are commonly attributed to the presence of a gap-edge vortex \citep{vandermarel2013, zhu2014, fuente2017}. 
Vortices may form due to Rossby wave instability (RWI) at the edge of a radial pressure bump in the disc. They are thought to act as dust traps \citep{barge1995}. %
The vortex is generally assumed to be induced by a planetary-mass companion internal to the vortex \citep[e.g. see ][]{vandermarel2013,fuente2017}, although the presence of dead zones in the disc have been shown to produce vortensity gradients that can trigger the formation of vortices via the RWI \citep{regaly2012,ruge2016}. In any case, very low values of disc viscosity are required for the RWI to be triggered and produce vortices \citep{regaly2012,ataiee13a}. Gas and dust simulations of vortices have shown that dust grains can strongly concentrate near the centre of the vortex, creating the asymmetries observed in IRS 48 and other discs \citep{birnstiel2013,fuente2017}. 

The appearance and stability of vortices depends strongly on the disc viscosity \citep{ataiee13a}, with a higher viscosity leading to shorter vortex lifetimes \citep{fu2014,zhu2014}, or no vortex formation at all. The effect of dust feedback on the gas can also suppress vortex formation, and disrupt vortices that may form in the disc \citep{inaba2006,fu2014b}, although recent 3D simulations have shown that this might not be the case \citep{lyra18a}. It has also been shown that vortex lifetimes are significantly reduced when simulations are initiated with lower-mass planets which undergo runaway gas accretion to reach their final mass \citep{hammer17a}. Despite these challenges, vortices are still the most commonly invoked mechanism for explaining high contrast ratio dust asymmetries in transition discs.

Vortices, however, are not the only pathway to generate asymmetries in the gas and dust distribution in a circumstellar disc. By dropping the requirement of low viscosity values, Jupiter mass planets have been shown to form eccentric cavities characterized by overdense features at the apocentre of the disc cavity, as a consequence of the clustering of eccentric orbits \citep{ataiee13a}; however, the density contrast ratio of these eccentric features is much lower than those found in observations. 

Another distinct mechanism is the presence of a (sub-)stellar companion that launches tidal streams of gas across the cavity, producing a pileup of material at its edge \citep{ragusa2017}. More massive companions were found to produce stronger asymmetries at the cavity edge. The important distinction between this mechanism, and a purely eccentric disc as explored by \cite{ataiee13a} is that the over-density is comoving with the gas around the cavity.
This feature has been observed in numerical simulations focusing on black hole circumbinary disc simulations \citep{farris2014, D'Orazio2016, ragusa2016, munoz2016, miranda2017}, and is seen to persist after several thousand orbits of the central binary.

The hypothesis of a binary-disc interaction was tested by \cite{price2018} for the transition disc HD142527 --- a disc which shows a strong azimuthal asymmetry in dust continuum emission \citep{casassus2013}. \cite{price2018} demonstrated that many of the morphological features of HD 142527 can be attributed to the observed companion \citep{biller2012,lacour2016}, and most importantly, showed that vortices are not the only contender in explaining dust asymmetries. 

Thus far, numerical simulations have demonstrated that binary companions can generate dust asymmetries with a contrast ratio of $\sim$10 \citep{ragusa2017, price2018}. However, transition discs have been observed with contrast ratios in far excess of this value. The most famous example is that of IRS 48, which has a contrast ratio greater than a factor of $\sim$100 \citep{vandermarel2013,vandermarel2015}.

In this paper we demonstrate that large dust grains can concentrate in the gas over-density that develops around high mass ratio companions on circular, co-planar orbits. This enhances the dust contrast in these systems. We investigate the kinematic signatures that this companion imprints on the circumbinary disc and show how these kinematic signatures may be used to distinguish between the vortex and circumbinary hypotheses in IRS 48.

\section{Observations of IRS 48}
\subsection{Gas structure}
Near infrared observations of IRS 48 by \citet{brown2012} revealed a ring-like structure of CO rovibrational emission approximately 30 au from the central star. More recent observations of CO isotopologue rotational lines taken with ALMA, tracing the colder gas, showed that CO is strongly depleted within this ring-like structure by at least a factor of 100 \citep{bruderer2014, vandermarel2016b}, and the $^{13}$CO 6--5 emission extends out to roughly 90 au \citep{vandermarel2016b}.

\subsection{Large dust grain asymmetry}
Large (sub-mm to cm) dust grains observed in IRS 48 show a high contrast ratio dust asymmetry \citep{vandermarel2013,vandermarel2015}, that peaks in intensity at approximately 60 au from the centre of the disk. 
The azimuthal extent of the dust emission also appears to change with dust grain size, with larger grains more azimuthally trapped than smaller grains \citep[][in prep]{vandermarel2015}. Any hypothesis that attempts to explain the nature of the dust trap in IRS 48 must reproduce this phenomenon.
\subsection{Ring-like dust features}

Ring-like features are observed across a range of wavelengths in the near infrared within and co-spatial with the large dust grain cavity. Beginning with the inner disk, there is evidence of dust emission from small grains and polycyclic aromatic hydrocarbons in a ring-like feature within the CO cavity \citep{geers2007,schworer2017,birchall2019}.

Further from the central star an East-West asymmetry has also been observed in 18.6 $\mu$m thermal emission \citep{geers2007,honda2018}. It was claimed by \cite{honda2018} that this thermal emission is not due to differences in the thermal structure between the East and West side of the disk. However, it is possible that there are temperature differences less than $\sim$ 45 K.

\section{Methods}\label{sec:meth}

We performed 3D smoothed particle hydrodynamical (SPH) simulations using the code {\sc phantom} \citep{phantom2018}. We explore the evolution of the gas disc, adding dust grains once the system has reached a quasi-steady state. The dust is coupled to the gas via aerodynamic drag, but we neglect the back reaction of the dust on the gas since the evolution of the gas disc is dominated by the gravitational interaction between the central star and companion. 

We model the dust particles using a two-fluid approach, where the gas and dust are modelled as two separate SPH populations \citep{twofluid2012}. This method is efficient for simulating dust grains with high Stokes number which are decoupled from the gas. This is an important difference between our dusty circumbinary disc simulations compared with those of \cite{ragusa2017}, who used a one-fluid approach assuming dust grains with low Stokes number \citep{laibe2014, price2015}.

In our dust simulations we model the dynamics of two grain populations with grain sizes $s_{\mathrm{grain}}$  of 100 $\mu$m and 1 mm, respectively, to study the concentration of dust grains in the pressure maximum. We assume a grain density $\rho_{\mathrm{grain}}=3$ g/cm$^3$. Since the Stokes number is the physically relevant parameter that governs the dust evolution, the simulation results would remain identical if we were to scale the size and density of the grains such that $s_{\mathrm{grain}} \rho_{\mathrm{grain}}$ remains constant \citep[see Eq. 3 of ][]{dipierro2015}. For this reason we refer to the 100 $\mu m$ and 1 mm as `medium' and `large' grains, respectively.

We used a separate simulation for each dust species. Since we neglect the back reaction of the dust onto the gas the final result would be the same if we simulated both grain sizes together. Separate simulations are necessary simply because simultaneous evolution of multiple types of dust particles is not yet available in the code.

We model the primary and secondary stars as sink particles \citep{bate1995}, which interact with each other and the gas and dust disc gravitationally. Particles can accrete onto both sink particles provided the SPH particles around the sink are gravitationally bound and within a specified accretion radius.

\subsection{Initial Conditions}
Our simulations contain a central protostar with a mass of 2 M$_{\odot}$, consistent with the mass estimated for IRS 48 \citep{brown2012}. We place a companion with a secondary-to-primary mass ratio $M_{\textrm{s}}/M_{\textrm{p}} = 0.2$, giving a companion mass of 0.4 M$_{\odot}$. We use accretion radii of 2 au and 0.5 au for the primary and companion stars respectively. These sink radii sizes mean that we do not resolve the accretion discs around each individual sink, but this is necessary to minimise computational time. The companion is initialised with a semi-major axis of 10 au and is initially on a circular, co-planar orbit. We discuss the consistency of a companion with this mass and semi-major axis in section \ref{sec:con}. The binary is free to evolve during the simulation, but given our small initial gas mass, the change in the binary orbital elements remains small.

We use $5 \times 10^6 $ SPH particles to model the gas disc, assuming a gas mass of $5\times 10^{-4}$ M$_{\odot}$, consistent with gas mass estimate of $5.5\times 10^{-4} M_\odot$ inside 90 au of IRS 48 \citep{vandermarel2016b}. We found in hindsight that increasing the gas mass by a factor of 5 was needed to reproduce the CO flux (see below). We discuss how this increase in gas mass remains consistent with the mass estimate by \citet{vandermarel2016b} in section \ref{sec:co}.
The gas is initially set up in Keplerian rotation in a co-planar annulus with the inner and outer radius as $R_\textrm{in} = 15$ au and $R_\textrm{out} = 40$ au, respectively. 
The radial extent of the gas in the disc dramatically increases as the simulations progress. 
We set the Shakura-Sunyaev alpha viscosity to $\alpha_{\textrm{SS}} \approx 1.5\times 10^{-3}$ by employing a constant SPH artificial viscosity parameter $\alpha_{\rm AV} = 0.11$.
The surface density of the gas is initialised as a power law with $\Sigma \propto R^{-p}$, with $p=1$.
The temperature profile of the disc is locally isothermal with $T(R) \propto (R/R_{\textrm{in}})^{-2q}$ and $q=0.25$.
The aspect ratio of the disc is determined by the $q$ index where $H/R = H_{\textrm{in}}/R_{\textrm{in}} (R/R_{\textrm{in}})^{\frac{1}{2}-q}$, and $ H_{\textrm{in}}=0.05$.

The gas disc is allowed to evolve and reach a quasi-steady state, taking on the order of $\sim$500 binary orbits to do so. To study the effects of dust grain migration and concentration in the dust trap, after 500 orbits we added dust grains with an initial density distribution set from the gas density distribution at that time assuming a dust to gas ratio of 1:100 for each grain size. Since we neglect the back reaction of the dust onto the gas, the amount of dust we add to the simulation is arbitrary.

We use a gas to dust particle ratio of 30:1 to prevent the dust particles from getting trapped under the gas particle smoothing length \citep{twofluid2012}. The gas+dust simulations are run for a further $\sim$30 orbits of the binary. This is a rather short period of time, but the addition of dust particles decreases the timestep size of the simulations drastically. This limits the duration for which we are able to track the evolution of the dust grains.

\subsection{Radiative transfer and synthetic observations}
We performed synthetic observations of our models using the Monte Carlo radiative transfer code {\sc mcfost} \citep{pinte06a,pinte2009}.

We employed a dust model based on the result of the SPH simulation by assuming a dust population with a power-law grain size distribution given by $dn/ds \propto s^{-m}$ between $s_{\min}$ to $s_{\max}$. We use the simulation that contains the medium sized dust grains since very little continuum emission comes from the much larger dust grains at the relevant wavelengths. The dust opacity was computed assuming spherical and homogeneous dust grains. There is evidence of grain growth in the dust trap of IRS 48 \citep{vandermarel2015}, so for our radiative transfer calculations we assume $s_{\min}=1\,\mu m$ and $s_{\max}=5\, mm$, with $m=3.0$. The total gas mass is taken from our SPH simulations. The dust mass is obtained by assuming a gas to dust ratio of 10, consistent with dust mass estimates in IRS 48 \citep{vandermarel2016}. We found that to reproduce the integrated $^{13}$CO emission, we needed to increase the gas mass by a factor of $\sim$5, so that the effective disc mass is $2.5 \times 10^{-3}$M$_\odot$, assuming a $^{13}$CO-to-H$_2$ abundance ratio of $2\times10^{-6}$. When producing our integrated $^{13}$CO emission images, we use this increased gas mass and use a gas to dust ratio of 10. The dust grain size and density is scaled such that the Stokes number of the grains remains unchanged when we scale the gas mass. Since the gas mass is small compared to the mass of the binary, the feedback of the gas onto the binary is negligible.

We assume that the dust and gas are in thermal equilibrium and the dust opacities are independent of temperature. We model the emission of the central sources as black bodies, assuming an effective temperatures of $T_{\textrm{p}}=10000$ K \citep{brown2012} and $T_{\textrm{s}}=3800$ K. % The stellar masses are set as the sink particle masses.

 We used $10^8$ Monte Carlo photon packets to compute the temperature and specific intensities. Images were then produced by ray-tracing the computed source function.

We assumed an inclination of $i=50^{\circ}$, a position angle of $\textrm{PA}=100^{\circ}$ \citep{bruderer2014}, and a source distance of 134 pc \citep{gaia2018} to create our simulated images of IRS 48. When comparing our simulated images with observations of IRS 48, we convolved our images with a Gaussian beam to match the beam size of the observations.

\section{Results}\label{sec:res}

\begin{figure*}
    \centering
    \includegraphics[width=0.8\linewidth]{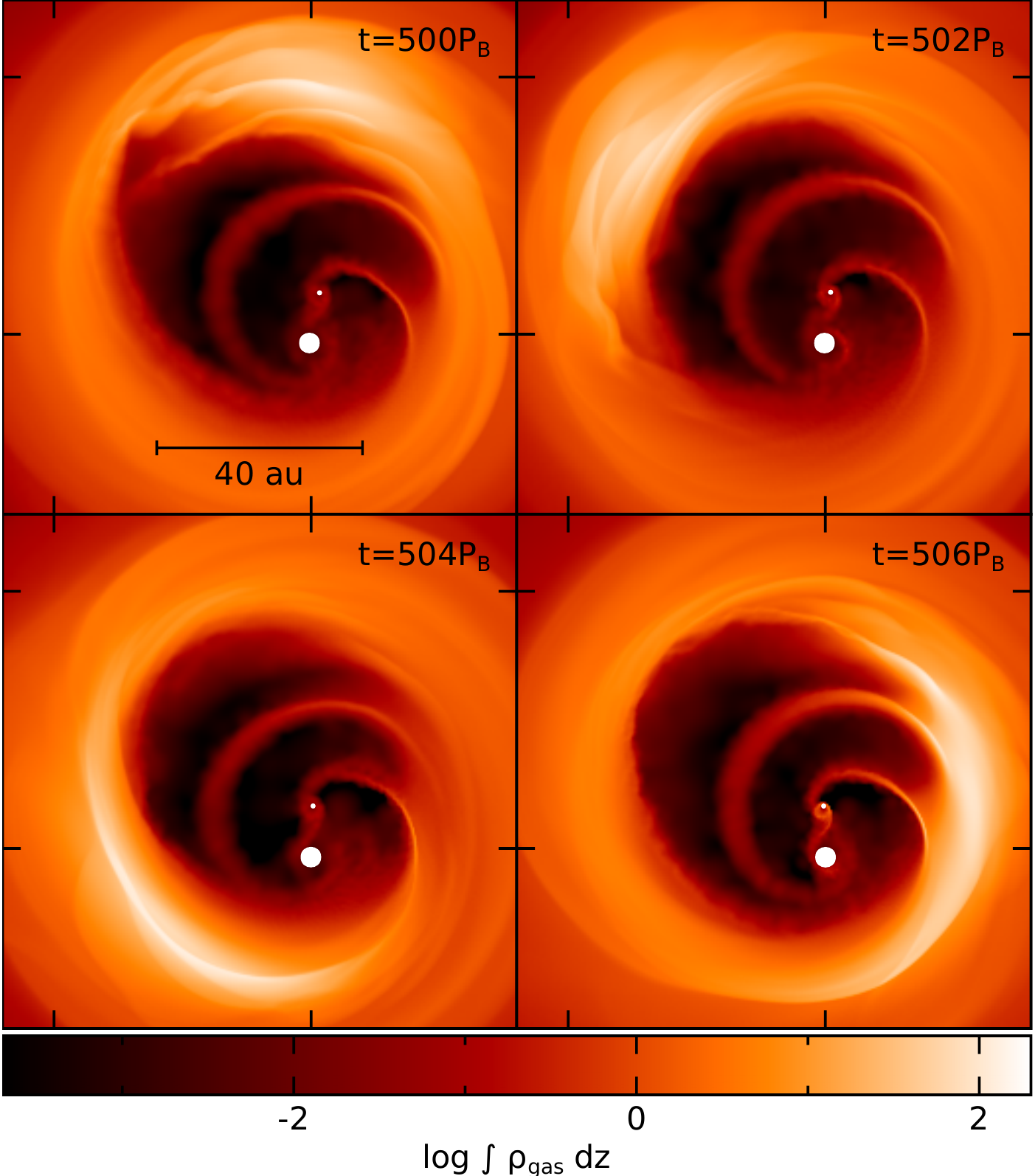}
    \vspace{-1em}
    \caption{Gas surface density of our circumbinary disc simulation from 500--506 binary orbits ($P_B$). The gas over-density orbits around the eccentric cavity at roughly the Keplerian velocity, and is not the result of a pile up of material at the apocentre of the eccentric disc, as in the case studied by \citet{ataiee13a}. Accretion flows on the binary stars appear as ring-like structures, and are generated every orbit of the binary companion. Outside of the cavity, the influence of the binary companion is still evident through the spiral density waves that are propagating through the disc. }
    \label{fig:orbit}
\end{figure*}

\begin{figure*}
    \centering
    \includegraphics[width=0.8\linewidth]{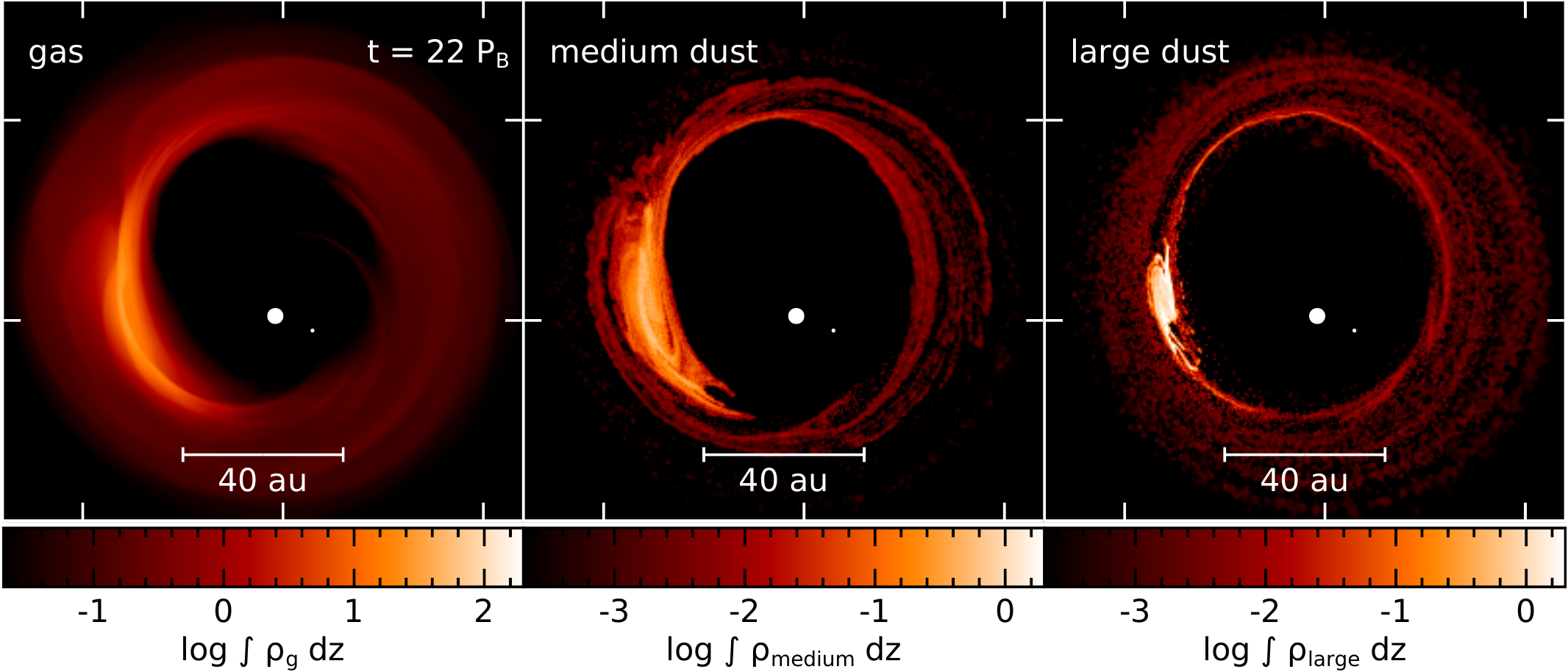}
    \vspace{-1em}
    \caption{Gas and dust column densities in our SPH simulations after $\sim$522 binary orbits. Dust was added after 500 binary orbits, once the eccentric cavity had reached a quasi steady-state, with dust density initially proportional to the gas surface density with a ratio of 1:100. The scale between the gas and the dust is shifted by a factor of 100. Therefore, we can directly see that the dust has concentrated if its surface density is elevated compared to that of the gas, which is clearly the case for the large dust grains. Note that since we are neglecting the back reaction of the dust onto the gas, the total amount of dust we add into the simulation is arbitrary. The dust to gas ratio seen in this Figure does not reflect the dust to gas ratio used when conducting the radiative transfer modelling.} \label{fig:sims}
\end{figure*}

Figure \ref{fig:orbit} shows the gas surface density after $\sim$500 binary orbits ($P_B$). The over-dense feature seen in the gas and dust is comoving with the fluid around the cavity, and is therefore different from the ``traffic jam'' of material that forms around an eccentric disc \citep{ataiee13a}. The over-density also lacks vortical motion \citep{ragusa2017}, and so is distinct from a vortex. The precise mechanism that leads to the formation of the cavity-edge over-density, its longevity and dependence on disc parameters, are not fully understood. We discovered that the instability leading to the cavity-edge over-density takes longer to initiate when the initial disc inner radius is large, but the instability does still form once the inner disc viscously evolves and reduces in radius. So although we used a close inner radius of $R_{\textrm{in}}=15$ au, it would not have made a large difference to the final outcome if we used a larger inner radius. We use a smaller inner radius since we spend less time evolving the system to reach the stage of developing the instability.

We show the outcome of adding dust into our simulations in Figure \ref{fig:sims}. Since the gas surface density changes by about an order of magnitude around the circumbinary disc, the Stokes number for each dust grain varies depending on its location in the disc. The medium sized dust grains (middle panel) have a Stokes number between 0.01---0.1, whereas the large dust grains have a Stokes number 0.1--1, depending if the grains are located within ot outside of the gas over-density.
Despite only being present in the simulation for $\sim$22 $P_B$, there is rapid migration and concentration in the gas pressure maximum for the large dust grains. Our main limitation preventing longer evolution time was the drastic increase in computation expense as the large dust grains concentrated. We were able to simulate the medium dust grains for a much longer period and observed an increased level of radial drift and concentration in the gas pressure maximum for these grains. We discuss this limitation more in Section \ref{sec:lim}.

It is evident in Figure \ref{fig:sims} that the dust grains have concentrated inside of the gas over-density. The initial dust distribution was set to be proportional to the surface density of the gas, with a gas to dust ratio of 100:1. The colorbar is scaled by this factor in Figure \ref{fig:sims} between the gas (left panel) and dust (middle and right panels). Therefore, we can see that the dust has concentrated if the surface density has increased, which is most obvious in the large dust grains.

\begin{figure}
    \centering
    \includegraphics[width=1\linewidth]{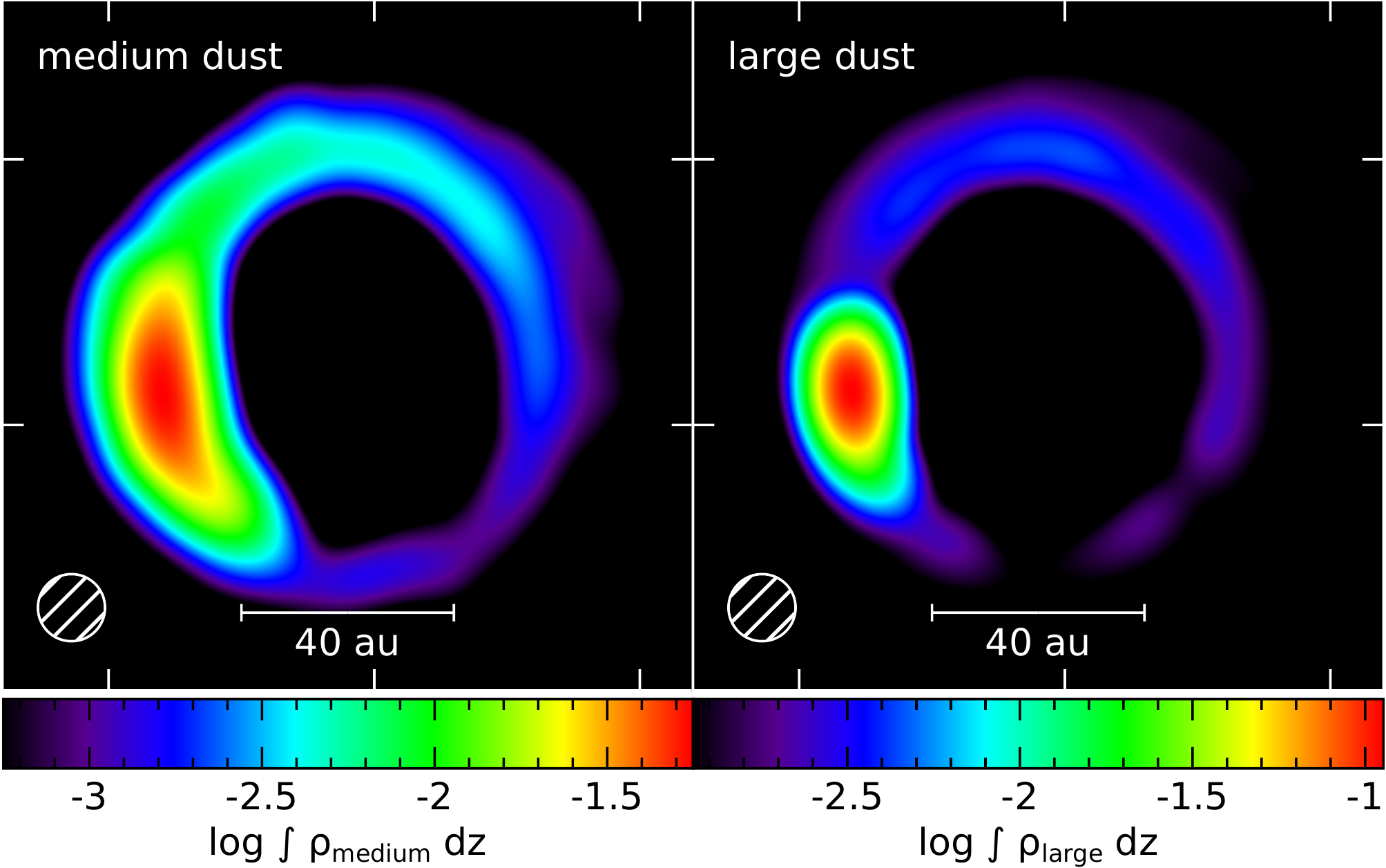}
    \vspace{-1em}
    \caption{Medium and large dust grain surface density distributions smoothed using a a beam size of 0.1" assuming the disc is observed at the distance of IRS 48. Dust grain simulations are the same as in middle and right panels of Figure \ref{fig:sims}. Medium sized grains concentrate inside the gas over-density, where a contrast ratio approaching 100:1 is achieved. Large dust grains concentrate more than the medium sized dust grains, exceeding a maximum contrast ratio of 100:1. The asymmetry in the medium dust grain surface density is more evident than in Figure~\ref{fig:sims}. Large dust grains display a more symmetric distribution. Similar asymmetry is seen in multi-wavelength observations of IRS 48 \citep[][in prep.]{vandermarel2015}.} 
    \label{fig:smooth}
\end{figure}

One revealing feature of Figure \ref{fig:sims} is the difference in azimuthal extent between the medium and large dust grains. The larger dust grains are more radially and azimuthally concentrated. This feature is in agreement with the more compact azimuthal distribution observed at longer wavelengths by \cite{vandermarel2015}. We show the dust grain surface density of our simulations smoothed to a beam size of 0.1" in Figure \ref{fig:smooth}, assuming the simulation is placed at the at a distance of 134 pc. This Figure makes it more clear that we are able to achieve a high dust contrast ratio approaching 100:1 in the medium sized dust grains, and exceeding 100:1 in the large dust grains.

Our simulated dust cavity is smaller ($\sim$20 au) than observed in IRS 48. The gas cavity that develops in a circumbinary disc is sensitive to the mass ratio, semi-major axis, eccentricity, and inclination of the companion \citep{Miranda&Lai2015,thun2017}. We did not explore a full range of possible orbits for the companion, but exploring this parameter space is a promising avenue to resolve this discrepancy. Given this, we scaled our code units prior to conducting the radiative transfer calculations to better match the observed cavity size in IRS 48. The code units are scaled such that the length unit was increased by 30\%, and the time unit scaled by an amount to match the change in length scale, according to equation 33 in \cite{phantom2018}. This means that the semi-major axis of our binary companion is scaled to 13 au.

\subsection{Dust Continuum}\label{sec:cont}

\begin{figure}
    \centering
    \vspace{-0.5em}
    \includegraphics[width=1.\linewidth]{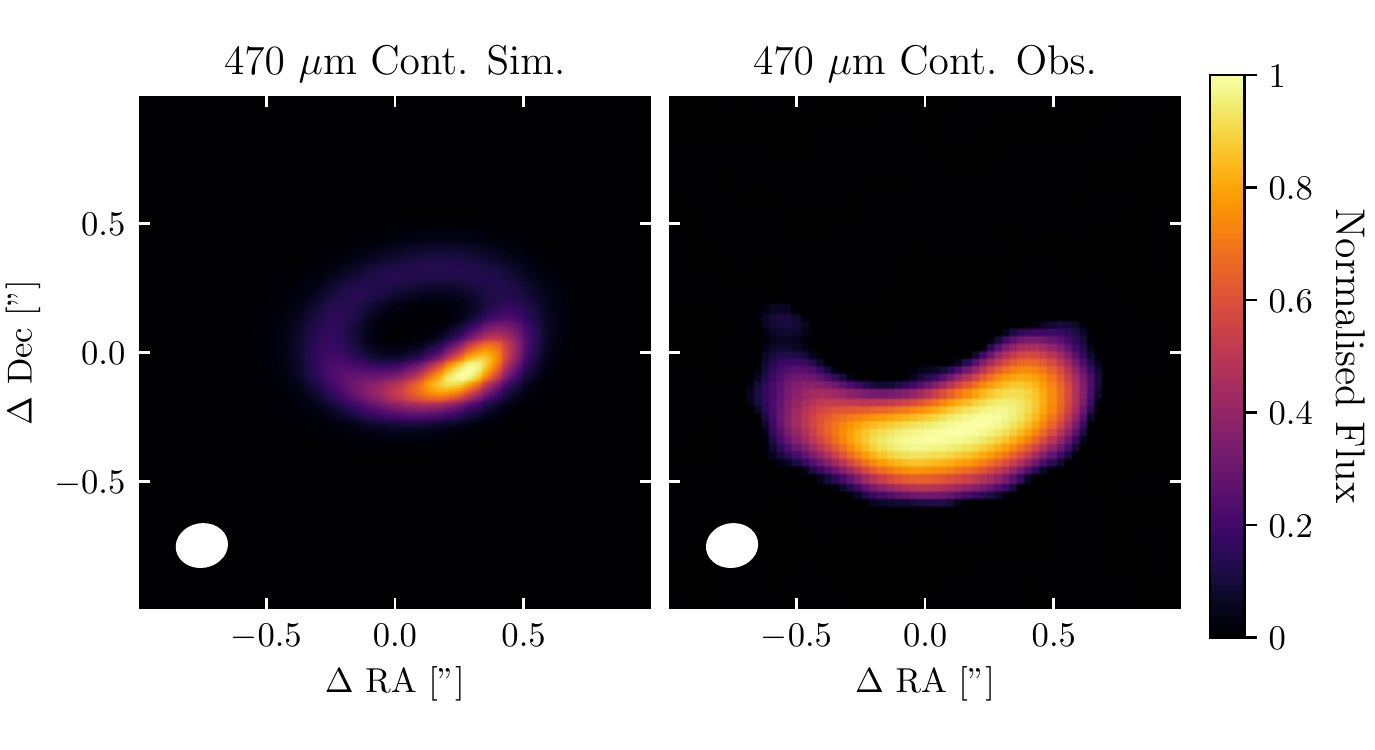}
    \vspace{-1em}
    \caption{Comparison of our synthetic 690 GHz continuum emission observations (left panel) with the 690 GHz continuum emission presented in \protect\cite{vandermarel2015,vandermarel2016b} (right panel).
	All scales are in normalised flux. The left panel shows the dust continuum emission when we match the kinematics of our simulation with that of IRS 48 (see Figure \ref{fig:velocity}). Our synthetic observations have been convolved with a Gaussian beam to match the observational data (white ellipse in the bottom left of each panel). Our simulated cavity size is smaller than that of IRS 48, however better agreement could be obtained by allowing the gas only simulation to advance for longer before adding in dust. The cavity size also depends on the orbit of the companion. Choosing a higher mass ratio for the companion (either by increasing the companion mass, or decreasing the primary mass), or increasing the eccentricity of the companion orbit, are two ways that we could increase the cavity size. We discuss this point further in section \ref{sec:lim}.}
	\label{fig:continuum}
\end{figure}

Figure \ref{fig:continuum} compares our synthetic observations (left panel) to the 690 GHz continuum emission presented in \cite{vandermarel2015,vandermarel2016b} (right panel). The medium sized dust grains have been allowed to evolve for $\sim 45P_B$ prior to generating the synthetic observations.
We recreate the asymmetric continuum emission observed in IRS 48 with modest agreement in density contrast and spatial distribution of the dust grains. Some notable differences between our simulated images, and the observed dust emission in IRS 48, are the radial distance and extent of the emission. We discuss ways in which we can increase the cavity size of our circumbinary disc in section \ref{sec:lim}. The radial extent of the dust trap in IRS 48 is resolved at the wavelength of the observations in Figure \ref{fig:continuum}. A substantial amount of the emission in these observations could be coming from smaller dust grains which have a lower Stokes number than the grains in our simulations. From our simulations containing the medium and large size dust grains, we can see that grains with Stokes numbers closer to unity tend to be more radially and azimuthally concentrated. Following the evolution of smaller dust grains in our simulations may relieve this discrepancy since they are expected to have a larger radial extent.

As the dust over-density orbits around the eccentric cavity, its observational appearance changes, closely following the changes in the gas over-density morphology seen in the panel of Figure \ref{fig:orbit}. The left panel of Figure \ref{fig:continuum} shows our best match to the observations of the right panel. In this simulated image, the dust grains are travelling towards the apastron of their orbit, and is in a similar position to the gas over-density in the bottom right panel of Figure \ref{fig:orbit}.

Although our synthetic observations in Figure \ref{fig:continuum} are not able to completely match the contrast ratio observed in IRS 48, we have plausible explanations for how this may be achieved.
Firstly, in our simulations we studied dust grains that are marginally coupled to the gas motion.
We observed rapid radial and azimuthal concentration of the dust into the gas over-density for the 1mm dust grains. If it were feasible to run these simulations for a longer period of time, we see no reason why this trend would not continue, provided the dust back-reaction on the gas remains negligible. We discuss this more in section \ref{sec:lim}.

It is also reasonable to expect that modelling dust growth and fragmentation would affect the dust contrast ratio for a given grain size. Dust growth and fragmentation is largely a function of the relative velocity of the dust grains \citep{testi2014}. Grains cease growing if their relative velocity exceeds the fragmentation barrier. In an eccentric disc the relative velocity of dust grains would likely be a function of position around their orbit of the cavity. This may lead to either growth or fragmentation of dust within the gas over-density of the circumbinary disc, depending on where it is on its orbit around the cavity. How this affects the appearance of the azimuthal asymmetry remains unknown.

\subsection{Integrated $^{13}$CO Emission}\label{sec:co}

\begin{figure}
    \centering
    \vspace{-0.5em}
    \includegraphics[width=1.\linewidth]{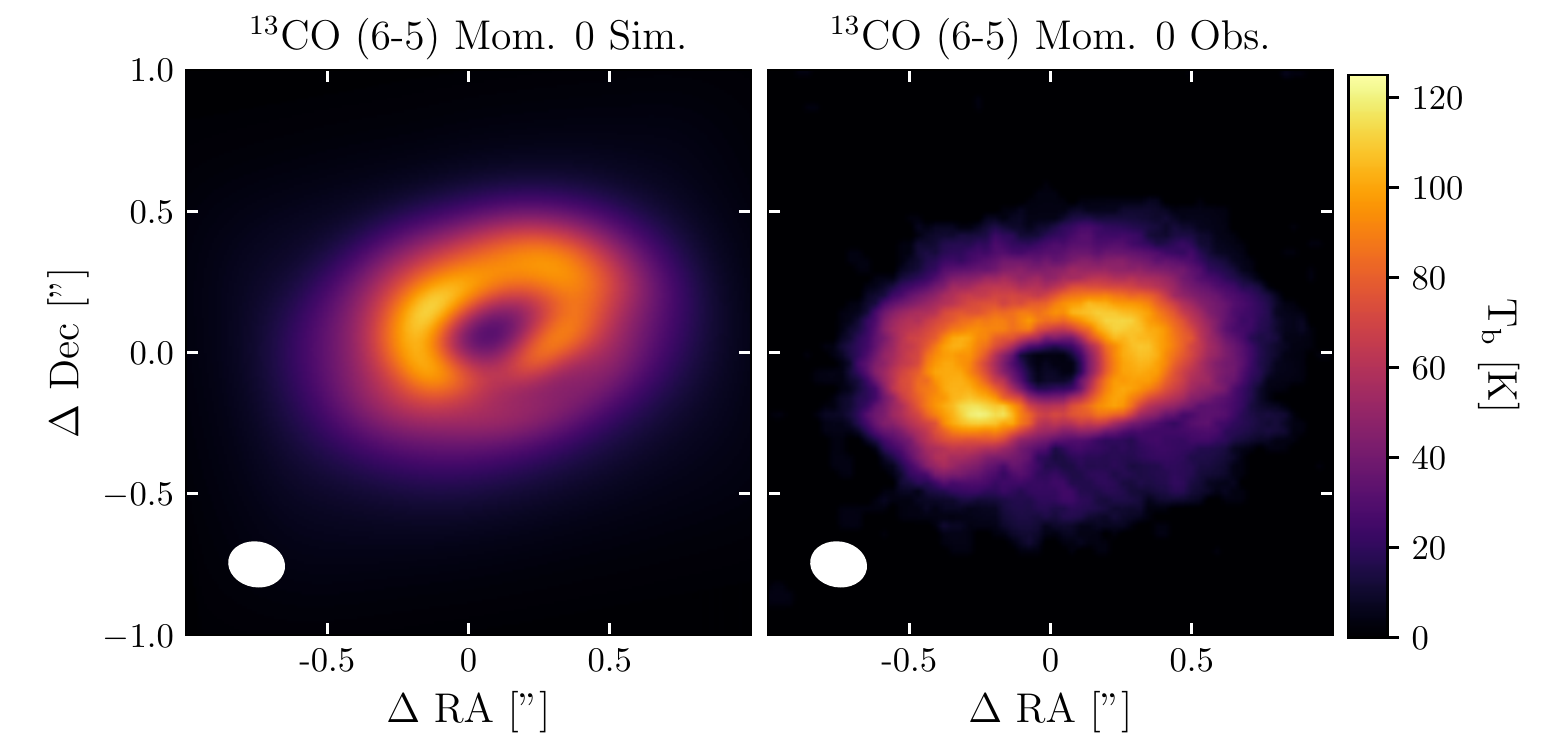}
    \vspace{-1em}
    \caption{Comparison of our $^{13}$CO (6-5) emission (left panel) with a comparison to the $^{13}$CO (6-5) emission of IRS 48 (right panel) shown in \protect\cite{vandermarel2016b}. The $^{13}$CO (6-5) emission in IRS 48 displays a strong drop in brightness temperature co-spatial with the dust trap. This is likely due to the $^{13}$CO emission being absorbed by the dust in the dust trap. We found that in order to reproduce this effect, a large amount of dust is required.}
	\label{fig:co}
\end{figure}

Figure \ref{fig:co} (left panel) shows the integrated $^{13}$CO 6--5 emission from our simulation, with a comparison to the integrated $^{13}$CO emission of IRS 48 presented in \cite{vandermarel2016b}. One feature worth discussing is the observed drop in $^{13}$CO brightness temperature co-located with the dust emission asymmetry. This feature is likely caused by the high concentration of dust grains at this location. The dust grains absorb the $^{13}$CO emission, reducing the brightness temperature in this region. This effect has two consequences.

First, the strong absorption of $^{13}$CO emission implies a very large concentration of dust at this location. In order for us to reproduce this effect in our radiative transfer calculations, we required a low gas-to-dust ratio of 10:1, consistent with the gas-to-dust ratio estimated by \cite{vandermarel2016b}. Note that this is a global gas-to-dust ratio, within the dust trap it is almost certainly much higher.
Such a vast amount of dust is required in this location that in our radiative transfer modelling, the dust over-density becomes optically thick. This means that the dust is actually more concentrated in the dust trap of IRS 48 than is implied by the dust emission observations. The level of $^{13}$CO emission absorption is also sensitive to the choice of the dust grain population distribution.

Second, a depletion in $^{13}$CO emission will reduce the estimated amount of $^{13}$CO in the disc. \cite{vandermarel2016b} estimated the disc mass of IRS 48 to be $\sim 0.5$M$_{\mathrm{J}}$, but this estimate would not account for the $^{13}$CO within the dust trap that is not being detected due to the absorption of $^{13}$CO emission by the dust grains. Therefore, the mass estimate by \cite{vandermarel2016b} really measures the mass contained outside of the dust trap.

In order for us to be able to reproduce the dust absorption of the $^{13}$CO, and match the brightness temperature of the $^{13}$CO, we needed to increase the gas mass of our simulations by a factor of $\sim 5$. The majority of the disc mass in our simulation is contained within the gas over-density (see left panel of Figure \ref{fig:sims}), and therefore this increase in gas mass is not inconsistent with the estimated disc mass outside of the dust trap by \cite{vandermarel2016b}.

We recreate the $^{13}$CO emission absorption feature in the synthetic image of our simulation in the left panel of Figure \ref{fig:co}. Since our dust distribution does not perfectly match that observed in IRS 48, the $^{13}$CO absorption feature is also not perfectly recreated. However we demonstrate that our model is capable of reproducing this feature.

We also note that the location of the $^{13}$CO emission absorption feature (and hence the dust over-density) is located outside of an inner $^{13}$CO ring. This is agreement with the observation of an inner CO ring observed in IRS 48 \citep{brown2012,vandermarel2016b}.

\subsection{Velocity Maps}

\begin{figure}
    \centering
    \vspace{-1em}
    \includegraphics[width=1\linewidth]{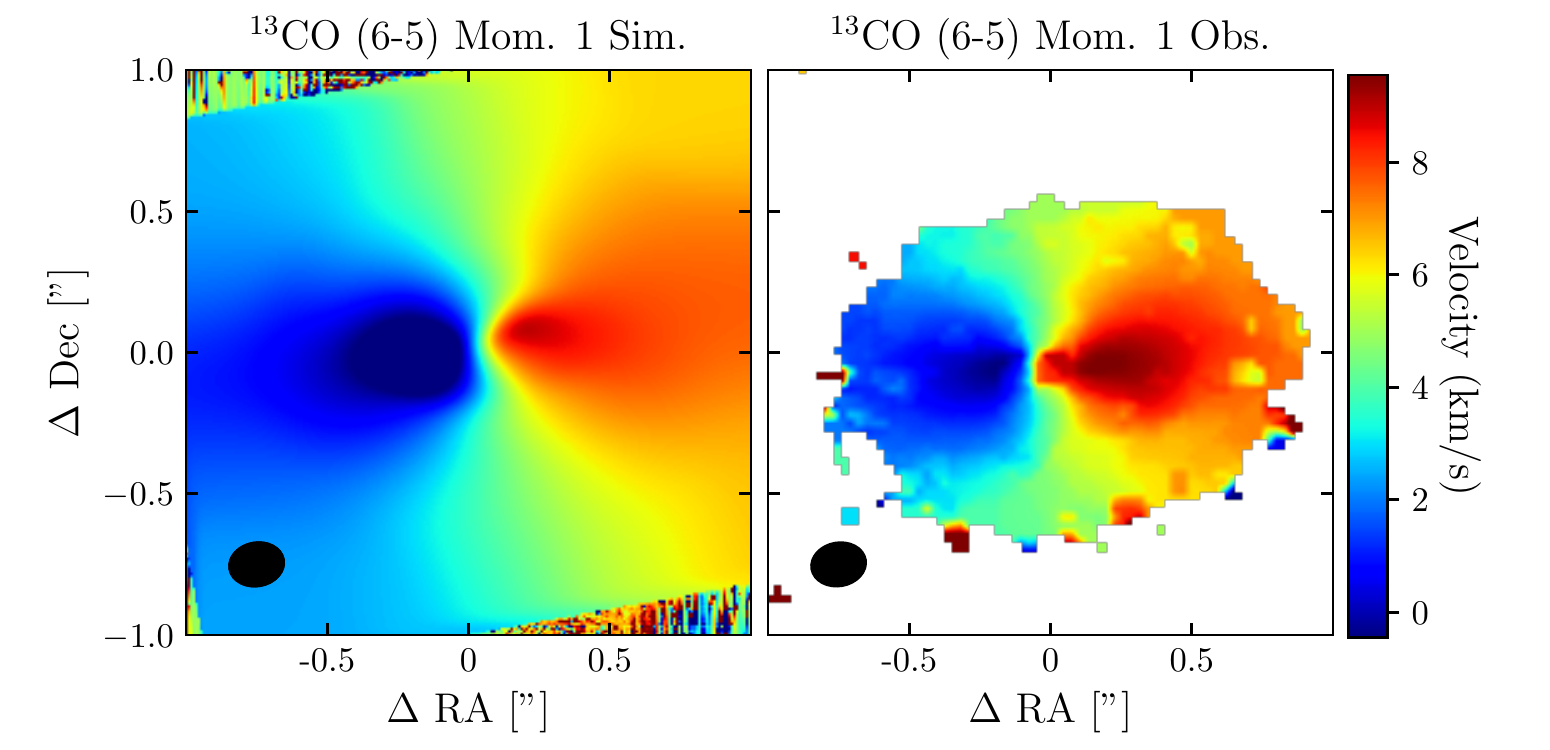}
    \vspace{-2em}
    \caption{Comparison of our synthetic velocity map (left panel) with the velocity $^{13}$CO (6--5) map of IRS 48 shown in \protect\cite{vandermarel2016b} (right panel). We convolve our synthetic map with a Gaussian beam matching the observational resolution (black ellipse in the bottom left of each panel). The kinematic map of IRS 48 shows asymmetry between the blue and red wings. In our synthetic map this is caused by the eccentric disc cavity and fast flowing gas inside the cavity. If the eccentricity vector of the eccentric disc was pointing along our line of sight, the kinematic map would show a more symmetric blue and red wing. As the eccentricity vector is offset to the line of sight in our synthetic map, an asymmetric blue and red wing is seen.}
    \label{fig:velocity}
\end{figure}

A peculiar asymmetry is observed in the kinematic map of $^{13}$CO (6--5) in IRS 48 (Figure \ref{fig:velocity}, right panel).
If this asymmetry is not an observational artefact, then it must be explained in models of vortices and eccentric discs alike.

The left panel of Figure~\ref{fig:velocity} shows our simulated kinematics. Our model naturally produces asymmetric kinematics due to the eccentric orbit of the gas. We also found that the position of the companion relative to the eccentric disc can leave an imprint in the velocity maps. This signal is much smaller than that of the eccentric disc itself, but may be observable in high spatial resolution molecular emission line observations. The asymmetry is observed from most azimuthal viewing angles, and becomes weaker when the eccentricity vector of the disc is pointing towards or away from the line of sight. In our model of IRS 48, the eccentricity vector of the disc is orientated nearly in the East-West direction. High velocity accretion flows onto the stars inside the cavity produces the central East-facing point in the red wing of the velocity maps in Figure \ref{fig:velocity}.
How exactly the kinematic maps change due to the orientation and inclination of the disc should be explored in future work.

With the sensitive nature of vortex formation in mind, it must be established whether or not they could form in discs that
produce the observed kinematics of IRS 48. Recent work \citep{perez2018, huang2018} demonstrates that simulated discs that contain vortices
show kinematic maps that are mostly symmetric, and the signature from the vortex is much smaller than what we predict for a circumbinary disc. For the vortex interpretation of the dust trap in IRS 48 to remain viable, it must be able to predict the asymmetry observed in the kinematic maps as well.

\subsection{Channel Maps}

\begin{figure*}
    \centering
    \includegraphics[width=1\textwidth]{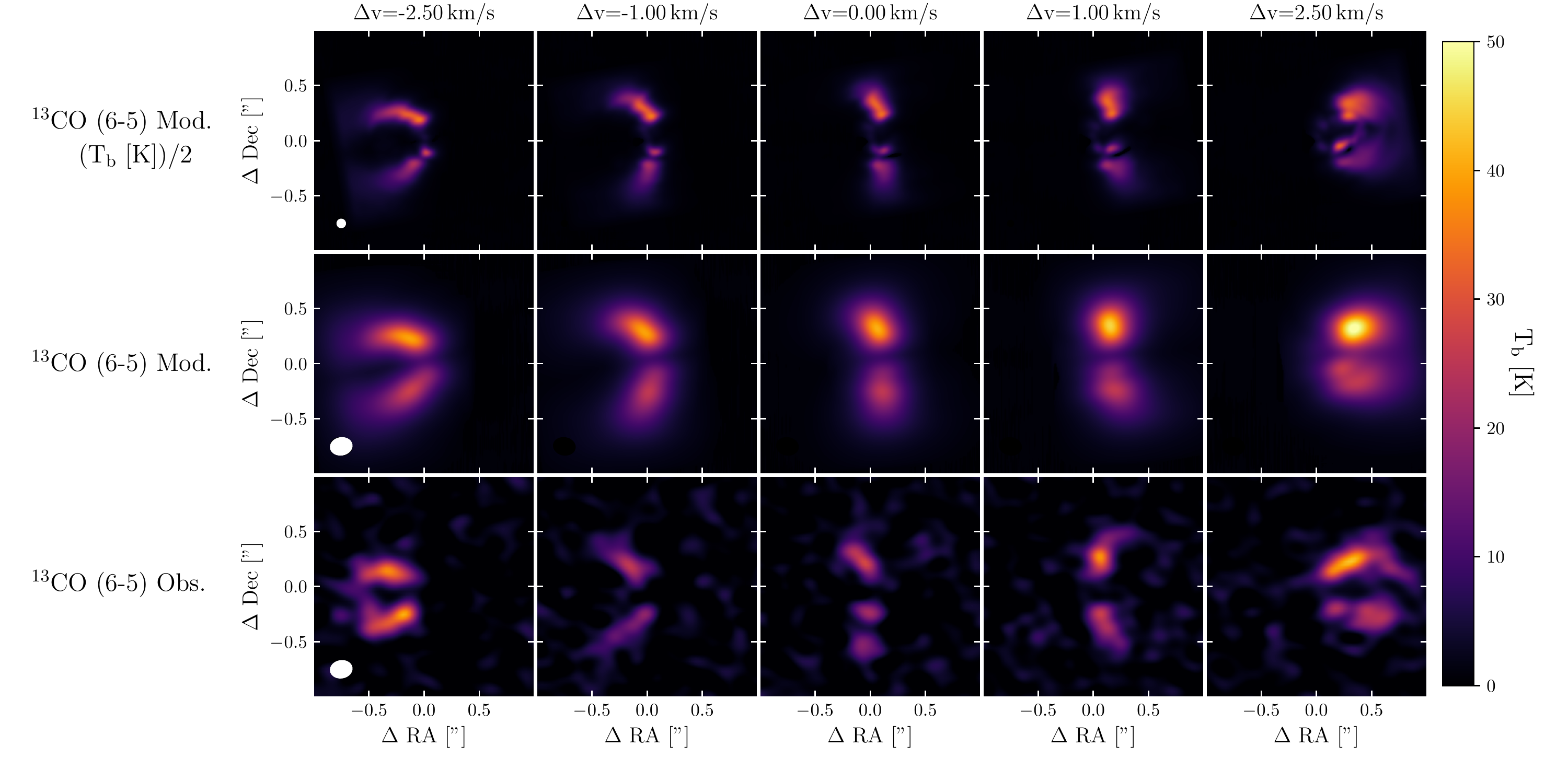}
    \caption{Comparison between our simulated channel maps convolved with two different beam sizes (top and middle rows), with a spectral resolution of 0.489 km/s, and the observed channel maps of IRS 48 (bottom row) presented in \protect\cite{vandermarel2016} for selected channels. Observed channel maps show deviations from Keplerian rotation. In particular, the angles at which the winglets in each channel propagate are not what is expected for a disk in circular rotation. For example, the rest frame channel of the system should have both winglets propagating at same angle as the position angle of the system ($10^{\circ}$ in this case), but it can be seen that they are not. The channel maps of our circumbinary disc simulation also show these features. The left and right wing are more compact in the observations than they are in our simulations, indicating our simulated total binary mass is higher than that in IRS 48, or that the inclination of the system is not as high as currently estimated (50$^{\circ}$).}
    \label{fig:channels}
\end{figure*}

Additional evidence for a circumbinary disc in IRS 48 is found in the individual channel maps of the $^{13}$CO (6--5) emission line observations presented in \cite{vandermarel2016b}. A companion of stellar mass will cause perturbations in the isovelocity contours of a disc. We also found that the gas over-density propagates spiral density waves as it orbits around the circumbinary disc, which can leave a distinct signature in the disc kinematics. In the middle panel of Figure \ref{fig:channels} we show channel maps from our circumbinary disc simulation at a spectral resolution of 0.489 km/s, and convolved with the same beam size as the observations of IRS 48 in the bottom panel. Even with the companion located at $\sim$ 10 au, its presence is discernible in the kinematics of the gas at the cavity edge and outer disc. 
The top panel of Figure \ref{fig:channels} shows the kinematics of our simulations as they would appear with a smaller beam size of 0.05". The `kinks' observed in the iso-velocity curves are being generated by some combination of the companion perturbing the disc, and the gas over-density propagating spiral density waves.

The most convincing evidence of the disc around IRS 48 being eccentric is the different angles at which the northern and southern winglets propagate, most noticeable in the channel that corresponds to the rest frame velocity of IRS 48 (middle column of Figure \ref{fig:channels}). If the disc around IRS 48 were in circular Keplerian rotation, both winglets would be pointing in an anti-parallel direction, and the angle at which they point would be equal to the position angle of the disc (estimated to be 10${^\circ}$). This is not the case here. One explanation for this is that the disc around IRS 48 is has an eccentric shape, and hence a velocity field that deviates strongly away from the velocity field of a disc with gas in a circular orbit.

Our simulation reproduces this feature due to the eccentricity that develops in a circumbinary disc. We found that rotating the simulation changes the angle at which the northern and southern winglets propagate. Placing the disc such that the eccentricity vector of the disc is pointing roughly along the East-West direction gives good agreement. Note that our simulated channel maps are the same as those used to produce the velocity map shown in Figure \ref{fig:velocity}, and also the $^{13}$CO 6-5 integrated emission in Figure \ref{fig:co}.

The blobby appearance in the southern wing of our simulation and in the observations arises due to the $^{13}$CO emission absorption discussed in section \ref{sec:co}.

The channel maps of our circumbinary disc simulation agree with the observed channel maps of IRS 48. The isovelocity curves in IRS 48 appear to be perturbed in a similar fashion to what is observed in our circumbinary disk model. Observing these features at higher spatial resolution would provide very strong evidence in favour of our circumbinary disc hypothesis of IRS 48.

Recent work by \cite{huang2018} made predictions for the imprint a vortex would leave on the channel maps and velocity map of IRS 48. The channel maps presented in their paper (their Figure 4) do not produce the peculiar structures seen in the channels maps of IRS 48. Furthermore, the velocity map presented (third row of their Figure 6) does not display an asymmetry between the blue and red-shifted sides of the disk. This does not exclude the possibility that these features can be produced by vortex models, but it shows that more work is required in understanding whether or not they are capable of being produced. The circumbinary disk model is able to reproduce both of these features.

One disagreement between our simulated channel maps and the observed channel maps of IRS 48 is the rate at which the left and right wings close, most noticeable in the left and right-most panels. This could be due to our sink particle masses being higher than the binary mass in IRS 48. We could achieve better agreement with the observed channel maps by reducing the mass of the two stars. This would not affect the development of the eccentric disc since this is sensitive to the mass ratio, and not the absolute mass, of the stellar companions \citep{Miranda&Lai2015}.

\section{Discussion}

\subsection{Consistency with Observations} \label{sec:con}
Binary searches in IRS 48 have resulted in no detections in the $\sim$18-800 au regime down to $m_K$=10.1 \citep{ratzka2005} and no detections in the $\sim$2.5-1400 au regime down to $m_K$=9.8 \citep{simon1995}. These limits were converted to mass estimates on the order of 100-150 M$_J$ \citep{wright2015}, below our companion mass of 0.4 M$_{\odot}$ at 13 au. However, IRS 48 is strongly extincted, making the detection limits highly uncertain and a massive companion may remain hidden \citep{wu2017,schworer2017}, potentially explaining why a direct detection has not yet been made. However, a recent companion search in IRS 48 has detected an asymmetry in the brightness profile of an inner dusty disc around IRS 48 at $\sim$15 au, a location consistent with the companion in our simulation \citep{birchall2019}.

Ring-like structures have been detected in the inner regions of the disc, close to our hypothetical stellar companion \citep{geers2007,brown2012,bruderer2014,birchall2019}. Accretion flows from the outer disc fall inside the cavity, producing ring-like structures, which can be seen in Figure \ref{fig:orbit}. In this Figure we show how the gas over-density at the cavity edge orbits at roughly the Keplerian velocity. The inner binary orbits on a shorter timescale, and the position and structure of the accreation flows inside the cavity vary on this timescale.

Furthermore, both the primary and companion stars will develop their own accretion discs. We do not resolve these structures in our simulations since they largely fall within the accretion radii of our sink particles, particularly for the primary star which has a larger sink radius. Studying the evolution and structure of the circumprimary and circumsecondary discs with the circumbinary disc is difficult with SPH due to the high numerical viscosity and low resolution present at low particle numbers.

\subsection{Numerical Limitations}\label{sec:lim}

In order to get a better agreement with the observed cavity size in IRS 48, we needed to increase the scale length of our simulations. Simulations of circumbinary discs by \cite{thun2017} indicate that a cavity is opened on a dynamical timescale of 10--20 orbits, but also that the eventual cavity size is set on the viscous timescale ($\sim$several thousand binary orbits). Given that we are running 3 dimensional simulations, it is challenging to run our simulations for a duration approaching the viscous timescale. We expect that our cavity would continue to increase in size and may eventually be consistent with the cavity size observed in IRS 48 \citep[see Figure 9 of][]{thun2017}, and the increase in scale length would no longer be necessary. Other works investigate the evolution of circumbinary discs over much longer timescales \citep[e.g.][]{miranda2017,thun2017}, however these tend to be done with two dimensional simulations which are cheaper.

Adding dust grains also increases the computational time of our simulations, since the timestep is now limited to be a fraction of the stopping time of the simulated dust species. Furthermore, the stopping time drops as the dust grains concentrate in the pressure maximum and around the cavity. The net result is that eventually we are no longer able to follow the evolution of the dust, particularly for the grains which migrate the fastest (e.g. the large grains in our simulation). As mentioned previously, we were able to simulate the medium dust grains for a longer duration, which was due to the fact that they took longer to concentrate and for a small portion to become trapped under the gas smoothing length.

Adjusting parameters such as the companion/primary mass ratio, the semi-major axis and eccentricity of the companion could also provide us with better agreement to the observed cavity size in IRS 48. However, performing parameter space searches with three-dimensional dusty simulations is difficult. Even without an extensive exploration of the parameter space, we can reproduce features of the dust trap such as contrast ratio and azimuthal trapping (Figures \ref{fig:sims} and \ref{fig:continuum}), the asymmetric velocity map (Figure \ref{fig:velocity}), and substructure in the channel maps (Figure \ref{fig:channels}). For this reason we believe that the simulation limitations discussed in this section do not substantially change our conclusion that IRS 48 is a circumbinary disc.

\section{Conclusions}

Our model of IRS 48 demonstrates that high contrast ratio asymmetries in dust continuum emission can be created in a circumbinary disc. Our simulations only include a stellar companion on a circular, co-planar orbit. Predictions of our circumbinary disc hypothesis are as follows:
\begin{enumerate}
\item We predict that there is a stellar companion in IRS 48 with an approximate mass of $\sim$ 0.4$M_\odot$ and with a semi-major axis of $\sim$ 10 au.
These values may differ if the companion is on an eccentric and/or inclined orbit, but for now we limit our analysis by assuming the companion to be on a nearly circular, co-planar orbit.
\item The interaction of this companion with the primary star causes the disc around the binary system to become eccentric, and an over-density to form which is co-moving with the gas around the cavity \citep{ragusa2017}. Owing to the eccentric shape of the disc, the position of the primary star should be offset from the projected centre of the gas disc.

\item Due to the non-Keplerian flow of the gas and dust in the eccentric disc, we predict that the velocity maps of IRS 48 will be asymmetric (see Figure \ref{fig:velocity}), with the degree of asymmetry growing with higher spatial resolution observations. We demonstrate that the isovelocity curves in the individual channel maps will display deviations and kinks (see Figure \ref{fig:channels}). The deviations should be more easily detected in higher spatial resolution observations.
\end{enumerate}

It is not clear how the dust grain distribution and velocity fields of the disc would differ with a companion on a more complicated orbit, although a first analysis was presented in \cite{price2018}.
This large parameter space of companion properties could be explored to yield a better match to observations of IRS 48. 

The mechanism for producing high contrast ratio dust horseshoes demonstrated in this paper has consequences that extend beyond IRS 48.
If it is shown that our proposed stellar companion in IRS 48 indeed exists, high contrast ratio dust horseshoes can no longer
be thought of as sign-posts for planets. Rather, they may be features of eccentric circumbinary discs around low mass ratio ($\gtrsim$0.2) binary stars. 

\section*{Acknowledgements}
We thank the anonymous referee for their useful comments which improved the quality of this manuscript.
J.C. acknowledges an Australian Government Research Training Program Scholarship.
C.P., D.J.P. and V.C. acknowledge funding from the Australian Research Council via
FT170100040, FT130100034, and DP180104235.
NC acknowledges funding from FONDECYT grant 3170680 and from CONICYT project Basal AFB-170002. 
E.R. and G.D. acknowledge funding from the European Research Council under the European Union's Horizon 2020 programme (grant No 681601). We used SPLASH \citep{price2007}.

%%%%%%%%%%%%%%%%%%%%%%%%%%%%%%%%%%%%%%%%%%%%%%%%%%

%%%%%%%%%%%%%%%%%%%% REFERENCES %%%%%%%%%%%%%%%%%%

% The best way to enter references is to use BibTeX:

\bibliographystyle{mnras}
\bibliography{paper} % if your bibtex file is called example.bib

%%%%%%%%%%%%%%%%%%%%%%%%%%%%%%%%%%%%%%%%%%%%%%%%%%

%%%%%%%%%%%%%%%%% APPENDICES %%%%%%%%%%%%%%%%%%%%%

% \appendix

%%%%%%%%%%%%%%%%%%%%%%%%%%%%%%%%%%%%%%%%%%%%%%%%%%

% Don't change these lines
\bsp	% typesetting comment
\label{lastpage}
\end{document}